\documentclass[12pt,preprint]{aastex}
\shorttitle{ABSORPTION OF 10-200 GeV GAMMA RAYS}

\shortauthors{Liu \& Bai}

\begin{document}

\title{ABSORPTION OF 10--200 GeV GAMMA RAYS BY RADIATION FROM BLR IN BLAZARS}

\author{H. T. Liu\altaffilmark{1,2} and J. M. Bai\altaffilmark{1}}
\email{liuhongtao1111@hotmail.com and baijinming@ynao.ac.cn}

\altaffiltext{1}{National Astronomical Observatories/Yunnan Astronomical
Observatory, the Chinese Academy of Sciences, Kunming, Yunnan
650011, P. R. China.}

\altaffiltext{2}{Graduate School of the Chinese Academy of
Sciences, Beijing, P. R. China; send offprint requests to
liuhongtao1111@hotmail.com}

\begin{abstract}

In this paper, we study the photon-photon pair production optical
depth for gamma-rays with energies from 10 to 200 GeV emitted by
powerful blazars due to the diffuse radiation field of broad line
region (BLR). There are four key parameters in the BLR model
employed to determine the $\gamma-\gamma$ attenuation optical
depth of these gamma-rays. They are the gamma-ray emitting radius
$R_{\gamma}$, the BLR luminosity $L_{\rm{BLR}}$, the BLR half
thickness $h$ and the ratio $\tau_{\rm{BLR}}/f_{\rm{cov}}$ of the
Thomson optical depth to the covering factor of BLR. For FSRQs, on
average, it is impossible for gamma-rays with energies from 10 to
200 GeV to escape from the diffuse radiation field of the BLR. If
$\it GLAST$ could detect these gamma-rays for most of FSRQs, the
gamma-ray emitting region is likely to be outside the cavity
formed by the BLR. Otherwise, the emitting region is likely to be
inside the BLR cavity. As examples, we estimate the photon-photon
absorption optical depth of gamma-rays with energies from 10 to
200 GeV for two powerful blazars, HFSRQ PKS 0405$-$123 and FSRQ 3C
279. Comparing our results with $\it GLAST$ observations in the
future could test whether the model employed and the relevant
assumptions in this paper are reliable and reasonable, and then
limit constraints on the position of the gamma-ray emitting region
relative to the BLR and the properties of the BLR.

\end{abstract}

\keywords{galaxies: active --- galaxies: jets --- gamma rays:
theory --- quasars: individual (3C 279, PKS 0405$-$123)}

\section{INTRODUCTION}
Thanks to the development of gamma-ray astronomy, the sample of
GeV or TeV gamma-ray emitters is increasing. With the detection of
high energy gamma-rays in over 60 blazars, such as 3C 279 and PKS
1510-089, in the GeV energy range by the EGRET experiment aboard
the Compton Gamma Ray Observatory (Catanese et al. 1997; Fichtel
et al. 1994; Lin et al. 1997; Mukherjee et al. 1997; Thompson et
al. 1995, 1996; Villata et al. 1997), an exceptional opportunity
presents itself for the understanding of the central engine
operating in blazars. Some of blazars have been also firmly
detected by atmospheric Cherenkov telescopes (ACTs) at energies
above 1 TeV, such as Mrk 421 (Punch et al. 1992) and Mrk 501
(Quinn et al. 1996). The Large Area Telescope instrument on the
Gamma-Ray Large Area Space Telescope ($\it GLAST$), the
next-generation high energy gamma-ray telescope which will be
launched next year, will observe gamma-rays with energies from 20
MeV to greater than 300 GeV, and have the unique capability to
detect thousands of gamma-ray blazars to redshifts of at least
$z=4$, with sufficient angular resolution to allow identification
of a large fraction of their optical counterparts (see e.g. Chen,
Reyes, \& Ritz 2004). The High Energy Stereoscopic System (HESS)
is an imaging atmospheric Cherenkov detector dedicated to the
ground based observation of gamma-rays at energies above 100 GeV
(Funk et al. 2004; Hinton 2004; Hofmann 2003), and successfully
detected TeV gamma-rays from many sources (see e.g. Aharonian et
al. 2005). The Major Atmospheric Gamma Imaging Cherenkov telescope
(MAGIC) is, currently the largest single-dish Imaging Air
Cherenkov Telescope in operation, the one designed to have the
lowest energy threshold of $\sim$ 30 GeV among the new Cherenkov
telescopes (see e.g. Baixeras et al. 2004). The MAGIC telescope
detected TeV gamma-rays from HESS J1813+178 (Albert et al. 2006).
$\it GLAST$, combined with new-generation TeV instruments such as
VERITAS, HESS and MAGIC, will tremendously improve blazar spectra
studies, filling in the band from 20 MeV to 10 TeV with high
significance data for hundreds of active galactic nuclei (AGNs)
(see Gehrels \& Michelson 1999). Future measurements of the
gamma-ray spectra shape and its variability of blazars would
tremendously improve our understanding to blazars.

Blazars are radio-loud AGNs characterized by luminous nonthermal
continuum emission extending from radio up to GeV/TeV energies,
very rapid variability, high and variable polarization, flat radio
spectrum, high brightness temperatures and supperluminal motion of
compact radio cores (e.g. Urry \& Padovani 1995). These unusual
characteristics of blazars are generally believed to be attributed
by emission from a relativistic jet oriented at a small angle to
the line of sight (Blandford \& Rees 1978). The spectral energy
distributions of blazars consist of two broad bumps. The first
component is from the synchrotron radiation processes. The second
component is contributed by inverse Compton emission of the same
electron population, including synchrotron self-Compton (SSC)
scattering synchrotron seed photons, and external Compton (EC)
scattering seed photons from sources outside the jet (see e.g.
B\"ottcher 1999). The external soft photon fields not only provide
target photons for EC processes to produce the second component,
but also absorb these gamma-rays from the EC processes, because
gamma-rays with energies above 10 GeV interact with optical--UV
photons and then could be attenuated by photon-photon pair
production collisions with the external soft photons. The
positions of gamma-ray emitting regions are still an open and
controversial issue in the researches on blazars. It is suggested
that gamma-rays are produced within broad line region (BLR) and
that the gamma-ray emitting radius $R_{\gamma}$ ranges roughly
between 0.03 and 0.3 pc (Ghisellini \& Madau 1996). It is argued
by Georganopoulos et al. (2001) that the radiative plasma in
relativistic jets of powerful blazars are within cavities formed
by BLRs. However, it is also argued by other researchers that the
gamma-ray emitting regions are outside BLRs (Lindfors, Valtaoja,
\& Turler 2005; Sokolov \& Marscher 2005).

Observations found a large population of X-ray strong FSRQs,
labeled HFSRQs by Perlman et al. (1998). HFSRQs have strong broad
emission lines, flat hard X-ray spectra similar to classical
FSRQs, but high synchrotron peak frequencies and steep soft X-ray
spectra similar to intermediate or high synchrotron peak frequency
BL Lac objects (e.g., Georganopoulos 2000; Padovani et al. 2003).
Therefore, it is expected that HFSRQs are likely to be GeV--TeV
gamma-ray emitters. For example, RGB J1629+4008 has a synchrotron
peak $\sim 2\times 10^{16}\/\ \rm{Hz}$ and a predicted Compton
flux $\nu F_{\rm{\nu}}$ of nearly $10^{-10}\/\ \rm{erg\/\
s^{-1}\/\ cm^{-2}}$ at $10^{25}\/\ \rm{Hz}$ (Padovani et al.
2002). However, the BLRs can reprocess and scatter the central UV
radiation to produce diffuse line emission and continuum
radiation, respectively. The two external diffuse radiation could
absorb gamma-rays with energies from 10 to 200 GeV. Then, it is
unknown whether these gamma-rays could be detected by $\it GLAST$
even if blazars intrinsically produce gamma-rays with energies
from 10 to 200 GeV in the observer's frame. In this paper, we
focus our efforts on the photon-photon absorption for these
gamma-rays by the two diffuse radiation fields of the BLRs in the
powerful blazars, and then hope to limit constraints on the
positions of gamma-ray emitting regions by comparing our results
with $\it GLAST$ observations in the near future.

The structure of this paper is as follows. Section 2 consists of
four subsections, in which subsection 2.1 presents the cross
section of pair production, subsection 2.2 the photon-photon
attenuation optical depth, subsection 2.3 the diffuse radiation
field of the BLR, and subsection 2.4 the dependence of
$\tau_{\gamma\gamma}$ on parameters. Section 3 presents the
photon-photon absorption optical depth for FSRQs, section 4
presents applications to individual sources, and the last section is for
discussions and conclusions. Throughout this paper, we use a flat
cosmology with a deceleration factor $q_0=0.5$ and a Hubble
constant $H_0=75 \/\ \rm{km \/\ s^{-1} \/\ Mpc^{-1}}$.

\section{THEORETICAL CALCULATION}
\subsection{The Cross Section of Pair Production}

Gamma-ray photons can be absorbed by the photo-annihilation/pair
creation process $\gamma+\gamma\rightarrow e^{+}+e^{-}$, for which
the total cross section is (e.g., Gould \& Schreder 1967; Jauch \&
Rohrlich 1976):
\begin{equation}
\sigma(x)=\frac{3\sigma_T}{16}(1-x^2)\left[(3-x^4)\ln\frac{1+x}{1-x}-2x(2-x^2)\right],
\end{equation}
where $\sigma_{\rm{T}}$ is the Thomson cross section, and $x$ is
the velocity of the pairs in the center of moment system and
defined as
\begin{equation}
x\equiv \sqrt{1-\frac{2}{\frac{h\nu_1}{m_{\rm{e}}
c^2}\frac{h\nu_2}{m_{\rm{e}}
c^2}(1-\cos{\theta})}}=\sqrt{1-\frac{2}{\varepsilon_1
\varepsilon_2 (1-\mu)}},
\end{equation}
where $h$ is the Plank constant, $m_{\rm{e}}$ the rest mass of
electron, $c$ the speed of light, $\varepsilon_1$
($\varepsilon_2$) is the energy of the high (low) energy photon in
units of the rest mass energy of electron, and $\mu$ is the cosine
of the collision angle $\theta$ (Coppi \& Blandford 1990). The
threshold condition of photon-photon pair creation process is
\begin{equation}
\frac{1}{2}\varepsilon_1\varepsilon_2(1-\cos{\theta})\ge 1.
\end{equation}
The total cross section reaches the maximum value with
$\sigma_{\rm{max}}\approx 0.26\sigma_{\rm{T}}$ at $x\approx 0.7$,
which corresponds to
\begin{equation}
\varepsilon_1\varepsilon_2(1-\cos{\theta})\approx 4.
\end{equation}
Then we have $\varepsilon_1\varepsilon_2\ga 2$ because $0\le
\theta \le \pi$. For the $\gamma-\gamma$ attenuation process, only
the photons with the product $\varepsilon_1\varepsilon_2\ga 2$
could have the maximum cross section $\sigma_{\rm{max}}\approx
0.26\sigma_{\rm{T}}$.

\subsection{The $\gamma-\gamma$ Attenuation Optical Depth}

The $\gamma-\gamma$ attenuation optical depth per unit length $dl$
at distance $R$ from the central black hole along the direction of
jet axis is
\begin{equation}
\frac{d}{dl}\tau_{\rm{\gamma\gamma}}(\varepsilon_1,R)=\int\int\int
\sigma(x)n_{\rm{ph}}(R,\varepsilon_2,\Omega)(1-\mu)d\varepsilon_2d\Omega,
\end{equation}
where $n_{\rm{ph}}(R,\varepsilon_2,\Omega)$ is the number density
of soft photon at the position $R$ per unit frequency range per
unit solid angle. For these gamma-rays produced at the position of
$R_{\rm{\gamma}}$ along the jet axis, the $\gamma-\gamma$
attenuation optical depth is
\begin{equation}
\tau_{\rm{\gamma\gamma}}(\varepsilon_1)=\int^{\infty}_{R_{\rm{\gamma}}}\frac{d}{dl}
\tau_{\rm{\gamma\gamma}}(\varepsilon_1,R)dR,
\end{equation}
and the escape probability of gamma photons of $\varepsilon_1$,
$P_{\rm{esc}}(\varepsilon_1)$, is
\begin{equation}
P_{\rm{esc}}(\varepsilon_1)=e^{-\tau_{\rm{\gamma\gamma}}(\varepsilon_1)}.
\end{equation}
In next paragraphs, we calculate the energy density distribution
of gamma-ray absorbing photons from both scattering and
reprocessing, and then estimate the $\gamma-\gamma$ attenuation
optical depth.

First, we calculate the intensity by integrating the emissivity
along lines of sight through BLR, which is assumed as a spherical
shell of clouds with the outer radius $r_{\rm{BLR,out}}$ and the
inner radius $r_{\rm{BLR,in}}$ (see Fig. 1). A fraction
$df_{\rm{cov}}(r)=n_{\rm{c}}(r)\sigma_{\rm{c}}(r)dr$ of the
central UV luminosity $L_{\rm{UV}}$ would then be reprocessed into
emission lines in $r\rightarrow r+dr$, where $n_{\rm{c}}(r)$ and
$\sigma_{\rm{c}}(r)$ are the number density and the cross section
of clouds at radius $r$, respectively. Since a thin spherical
shell has volume $4\pi r^2dr$, the emissivity (in $\rm{erg\/\
s^{-1}\/\ cm^{-3}\/\ ster^{-1}}$) of reprocessed radiation inside
the BLR at the radius $r$ is
\begin{equation}
j_{\rm{line}}(r)=\frac{L_{\rm{UV}}df_{\rm{cov}}}{16\pi^2 r^2
dr}=\frac{L_{\rm{UV}}n_{\rm{c}}\sigma_{\rm{c}}}{16\pi^2 r^2},
\end{equation}
where $\sigma_{\rm{c}}=\pi R_{\rm{c}}^{2}$ and $R_{\rm{c}}$ is the
radius of clouds at the radius $r$. For
$n_{\rm{c}}=n_{\rm{c0}}(r/r_{\rm{BLR,in}})^{-p}$ and
$R_{\rm{c}}=R_{\rm{c0}}(r/r_{\rm{BLR,in}})^q$, where $n_{\rm{c0}}$
and $R_{\rm{c0}}$ is the number density and the radius of clouds
at $r_{\rm{BLR,in}}$, respectively, the covering factor
$f_{\rm{cov}}$ of the entire BLR to the central UV radiation is
\begin{equation}
f_{\rm{cov}}=\int^{r_{\rm{BLR,out}}}_{r_{\rm{BLR,in}}}n_{\rm{c0}}\pi
R^2_{\rm{c0}}\left(\frac{r}{r_{\rm{BLR,in}}}\right)^{2q-p}dr.
\end{equation}
In this paper, we adopt the power law exponents preferred by Kaspi
\& Netzer (1999), $q=1/3$ and $p=1.5$. The central UV radiation is
also scattered by the BLR clouds into the diffuse continuum, and
the diffuse soft photons can absorb gamma-rays with energies above
10 GeV. A fraction $d\tau_{\rm{BLR}}=n_{\rm{e}}\sigma_{\rm{T}}dr$
of the central UV luminosity would then be scattered in
$r\rightarrow r+dr$, where $n_{\rm{e}}$ is the number density of
electrons in the BLR clouds at the radius $r$. Since this
spherical shell has volume $4\pi r^2 dr$, the emissivity of
scattered radiation inside the BLR at the radius $r$ is
\begin{equation}
j_{\rm{cont}}(r)=\frac{L_{\rm{UV}}d\tau_{\rm{BLR}}}{16\pi^2 r^2
dr}=\frac{L_{\rm{UV}}n_{\rm{e}}\sigma_{\rm{T}}}{16\pi^2 r^2},
\end{equation}
if the Thomson optical depth of the BLR $\tau_{\rm{BLR}}\ll 1$.
For $n_{\rm{e}}=n_{\rm{e0}}(r/r_{\rm{BLR,in}})^{-s}$, where
$n_{\rm{e0}}$ is the number density of electrons in the BLR clouds
at $r_{\rm{BLR,in}}$, the Thomson optical depth of the entire BLR
to the central UV radiation is
\begin{equation}
\tau_{\rm{BLR}}=\int^{r_{\rm{BLR,out}}}_{r_{\rm{BLR,in}}}n_{\rm{e0}}
\sigma_{\rm{T}} \left(\frac{r}{r_{\rm{BLR,in}}}\right)^{-s}dr.
\end{equation}
In this paper, we adopt the power law exponent preferred by Kaspi
\& Netzer (1999), $s=1$. From equations (8) and (9) and the
relation of the luminosity of broad emission lines from the BLR
and the central UV luminosity
$L_{\rm{BLR}}=f_{\rm{cov}}L_{\rm{UV}}$ (D'Elia et al. 2003), we
get the expression for the diffuse line emission
\begin{equation}
j_{\rm{line}}(r)=\frac{L_{\rm{BLR}}r^{2q-p-2}}{16\pi^2\int_{r_{\rm{BLR,in}}}^{r_{\rm{BLR,out}}}r^{2q-p}dr}.
\end{equation}
From equations (10) and (11) and
$L_{\rm{BLR}}=f_{\rm{cov}}L_{\rm{UV}}$, we also get the emissivity
for the diffuse continuum radiation
\begin{equation}
j_{\rm{cont}}(r)=\frac{L_{\rm{BLR}}\tau_{\rm{BLR}}r^{-s-2}}{16\pi^2
f_{\rm{cov}}\int_{r_{\rm{BLR,in}}}^{r_{\rm{BLR,out}}}r^{-s}dr}.
\end{equation}

The intensity of radiation with angle $\theta$ to the jet axis at
a position $R$ is given by
\begin{equation}
I(R,\theta)=\int^{l_{\rm{max}}}_{0}j(r)dl,
\end{equation}
where $j(r)$ represents $j_{\rm{cont}}(r)$, $j_{\rm{line}}(r)$ or
$j_{\rm{cont}}(r)+j_{\rm{line}}(r)$, and vanishes when
$r>r_{\rm{BLR,out}}$ or $r<r_{BLR,in}$. We can get the following
relations from the triangle in Figure 1
\begin{equation}
r^2=R^2-2Rl\cos\theta+l^2,
\end{equation}
\begin{equation}
l_{\rm{max}}=R\cos\theta+R\sqrt{\cos^2\theta+\left(\frac{r_{\rm{BLR,out}}}{R}\right)^2-1}.
\end{equation}
The energy density of radiation from the BLR is expressed as
\begin{equation}
U(R)=\int\int \frac{I(R,\theta)}{c}d\Omega=\frac{2\pi}{c}
\int^{\pi}_{0} I(R,\theta)\sin \theta d\theta,
\end{equation}
and is a function of distance $R$ from the center along the
direction of the jet axis.

\subsection{The Diffuse Radiation Field of BLR}

Though, the emission lines $Ly\alpha$, $C IV$, $Mg II$,
$H_{\gamma}$, $H_{\beta}$ and $H_{\alpha}$ are the major
contributors to the total broad line emission, the contributions
from all the emission lines reported by Francis et al. (1991) and
the emission line $H_{\rm{\alpha}}$ reported by Gaskell et al.
(1981) are included to estimate the total flux of broad emission
lines, $F_{\rm{BLR}}$. Together 35 components of emission lines
are considered to estimate the total flux of broad emission lines.
The sum of line ratios $N_{\rm{\nu}}$, relative to the $Ly\alpha$
with a reference value of $N_{\rm{\nu}}=100$, of all the 35
components of emission lines is equal to 555.77 (Francis et al.
1991). The relative intensities of the 35 components of emission
lines are presented in Figure 3. It is assumed that the flux of a
certain emission line is equal to
\begin{equation}
F_{\rm{\nu}}=\frac{N_{\rm{\nu}}}{555.77}F_{\rm{BLR}}.
\end{equation}
According to the definition of flux, we have relations of
$F_{\rm{BLR}}=\int I_{\rm{line}}(R,\theta)\cos\theta d\Omega$ and
$F_{\rm{\nu}}=\int I(R,\nu,\theta)\cos\theta d\Omega$. Comparing
the two relations with equation (18), we have a relation of
$I(R,\nu,\theta)=N_{\rm{\nu}}I_{\rm{line}}(R,\theta)/555.77$
between the total intensity $I_{\rm{line}}(R,\theta)$ and the
monochromatic intensity $I(R,\nu,\theta)$ for the emission line.
Then the photon number density of the emission line along the
direction of $\theta$ at the position of $R$ is
\begin{equation}
n_{\rm{ph}}(R,\nu,\Omega)=\frac{I(R,\nu,\theta)}{h\nu
c}=\frac{N_{\rm{\nu}}}{555.77}\frac{I_{\rm{line}}(R,\theta)}{h\nu
c}.
\end{equation}
A diluted blackbody spectrum at the position of $R$ along the
direction of $\theta$ is assumed for the diffuse continuum as
\begin{equation}
n_{\rm{ph}}(R,\nu,\Omega)=\frac{2n(R,\theta)\nu^2}{c^3\left[\exp\left(\frac{h\nu}{kT}\right)-1\right]}
=n(R,\theta)n_{\rm{bb}}(\nu,T),
\end{equation}
where $k$ is the Boltzmann constant, and $n(R,\theta)$ is the
normalization factor to a blackbody spectrum $n_{\rm{bb}}(\nu,T)$
and can be calculated by the formula
\begin{equation}
n(R,\theta)=\frac{I_{\rm{cont}}(R,\theta)}{c\int^{\nu^U_2}_{\nu^L_2}n_{\rm{bb}}(\nu,T)h\nu
d\nu},
\end{equation}
because $I_{\rm{cont}}(R,\theta)=cn(R,\theta)\int
n_{\rm{bb}}(\nu,T)h\nu d\nu$. In the calculations of the
$\gamma-\gamma$ attenuation optical depth, the soft photon
frequency is considered in the range from $\nu^L_2=10^{14.6}\/\
\rm{Hz}$ to $\nu^U_2=10^{16.5}\/\ \rm{Hz}$ for the diffuse
continuum at the bands of the optical to UV, and the blackbody
temperature $T=10^5$ K is assumed for the inner regions of the
accretion disk.

\subsection{The Dependence of $\tau_{\gamma\gamma}$ on Parameters}

According to the interpretation of the data obtained in many
reverberation campaigns, production of broad emission lines in
quasars is peaked around some distance (see Peterson 1993; Kaspi
2000; Sulentic et al. 2000). Wu et al. (2004) derived an empirical
relation between the BLR size $r_{\rm{BLR}}$ deduced from the
reverberation mapping and $H_{\beta}$ luminosity
\begin{equation}
\log r_{\rm{BLR}}=-1.695+0.684 \log
\frac{L_{H_{\beta}}}{10^{42}\/\ \rm{erg \/\ s^{-1}}} \/\ \rm{pc}.
\end{equation}
From equation (18) and the line ratio of $N_{\rm{\nu}}=22$ for the
emission line $H_{\beta}$, the empirical relation is also
expressed as
\begin{equation}
r_{\rm{BLR}}=0.250\times \left(\frac{L_{\rm{BLR}}}{10^{45}\/\
\rm{erg \/\ s^{-1}}}\right)^{0.684} \/\
\rm{pc}=0.250L_{\rm{BLR,45}}^{0.684} \/\ \rm{pc},
\end{equation}
where $L_{\rm{BLR,45}}=L_{\rm{BLR}}/(10^{45}\/\ \rm{erg \/\
s^{-1}})$. We assume that the distance $r_{\rm{BLR}}$ deduced from
equation (23) represents the median of $r_{\rm{BLR,in}}$ and
$r_{\rm{BLR,out}}$, that is, $r_{\rm{BLR,in}}=r_{\rm{BLR}}-h$ and
$r_{\rm{BLR,out}}=r_{\rm{BLR}}+h$, where $h$ is the half thickness
of the spherical shell of clouds.

We derived an average broad line luminosity of
$L_{\rm{BLR}}=10^{45.35}\/\ \rm{erg\/\ s^{-1}}$ for 25 FSRQs from
Cao \& Jiang (1999), and took it as the typical value of
$L_{\rm{BLR}}$ for powerful blazars. The typical value of
$L_{\rm{BLR}}=10^{45.35}\/\ \rm{erg\/\ s^{-1}}$ for FSRQs
corresponds to a distance $r_{\rm{BLR}}\simeq 0.43 \/\ \rm{pc}$.
In order to illustrate the radial dependence of the energy density
of the diffuse radiation on the two parameters $r_{\rm{BLR}}$ and
$h$, equivalently $r_{\rm{BLR,in}}$ and $r_{\rm{BLR,out}}$, we
firstly consider four cases with four values of $h$ at a fixed
$r_{\rm{BLR}}=0.43 \/\ \rm{pc}$, and then consider four cases with
four values of $r_{\rm{BLR}}$ at a fixed $h=0.20 \/\ \rm{pc}$ by
assuming a ratio $\tau_{\rm{BLR}}/f_{\rm{cov}}=1$. The relevant
energy density distributions of the diffuse radiation fields are
plotted in Figure 2. It can be seen that for each case the energy
density peaks around the inner radius $r_{\rm{BLR,in}}$, and
decreases rapidly beyond the outer radius $r_{\rm{BLR,out}}$. On
the whole, the energy density inside the inner radius increases as
the parameter $h$ increases for a fixed $r_{\rm{BLR}}$, and the
average energy density inside the outer radius also increases as
$h$ increases. The energy density inside the inner radius
decreases as the parameter $r_{\rm{BLR}}$ increases for a fixed
$h$, and the average inside the outer radius also decreases as
$r_{\rm{BLR}}$ increases. The energy density outside the inner
radius does not monotonously increase with $h$ for a fixed
$r_{\rm{BLR}}$ (see Fig. 2c). At a fixed $r_{\rm{BLR}}$, the
increasing of $h$ results in the increasing of $r_{\rm{BLR,out}}$
and decreasing of $r_{\rm{BLR,in}}$. It can be seen from Figure 2c
that the variation of the inner radius can significantly change
the energy density distributions, but the energy density
distributions slightly alter as the outer radius changes. At a
fixed $h$, the increasing of $r_{\rm{BLR}}$ results in the
increasing of $r_{\rm{BLR,in}}$ and $r_{\rm{BLr,out}}$. It is
obvious that the energy density tends to become smaller as either
$r_{\rm{BLR,in}}$ or $r_{\rm{BLR,out}}$ becomes larger (see Fig.
2a, b). The energy density outside the inner radius does not
monotonously decrease with $r_{\rm{BLR}}$ for a fixed $h$ (see
Fig. 2a, b). In addition, the radial dependencies of the energy
densities have a very similar behavior as $R$ is large enough. In
order to study the dependence of $\tau_{\gamma\gamma}$ on
parameters $h$, $R_{\gamma}$ and $r_{\rm{BLR}}$, a ratio of
$\tau_{\rm{BLR}}/f_{\rm{cov}}=1$ is assumed and the BLR luminosity
of $L_{\rm{BLR}}=10^{45.35}\/\ \rm{erg\/\ s^{-1}}$ is adopted for
FSRQs.

We assume $R_{\gamma}=r_{\rm{BLR}}$ and take four values of $h$ to
study the dependence of the photon-photon absorption optical depth
on the parameter $h$. The calculated results are presented in
Figure 4. The parameter $h$ influences the photon-photon
absorption optical depth, but the influence of $h$ on the
calculated results is not very important (see Fig. 4a). It can be
seen from Figure 4a that the optical depth $\tau_{\gamma\gamma}$
does not monotonously increase with $h$, qualitatively changing
the dependence on $h$ with $E_{\rm{\gamma}}$. We study the
dependence of $\tau_{\gamma\gamma}$ on $h$ for four values of
$E_{\gamma}$ to check if this effect is real. The calculated
results are presented in Figure 4b, and we can see from Figure 4b
that the optical depth $\tau_{\gamma\gamma}$ does not monotonously
increase with $h$, qualitatively changing the dependence on $h$
with $E_{\rm{\gamma}}$. To better show this effect, we normalize
$\tau_{\gamma\gamma}(h)$ by the factor $\tau_{\gamma\gamma}(0.05
\/\ \rm{pc})$ for the four curves in Figure 4b, and the normalized
results are presented in Figure 4c. For the curves I, II and III
with the relevant $E_{\gamma}$ at 50, 100 and 150 GeV, the optical
depth $\tau_{\gamma\gamma}$ at four $h$ have an order of
$\tau_{\gamma\gamma}(0.05 \/\ \rm{pc})< \tau_{\gamma\gamma}(0.30
\/\ \rm{pc})< \tau_{\gamma\gamma}(0.20 \/\ \rm{pc})<
\tau_{\gamma\gamma}(0.15 \/\ \rm{pc})$ (see Fig. 4c). For the
curve I+II with $E_{\gamma}$ at 200 GeV, the optical depth
$\tau_{\gamma\gamma}$ at four $h$ have a sequence of
$\tau_{\gamma\gamma}(0.30 \/\ \rm{pc})< \tau_{\gamma\gamma}(0.05
\/\ \rm{pc})< \tau_{\gamma\gamma}(0.20 \/\ \rm{pc})<
\tau_{\gamma\gamma}(0.15 \/\ \rm{pc})$ (see Fig. 4c). It is
obvious that the optical depth $\tau_{\gamma\gamma}$ does not
monotonously increase with $h$ and the dependence of
$\tau_{\gamma\gamma}$ on $h$ qualitatively changes with
$E_{\rm{\gamma}}$ (see Fig. 4c). Then this effect is real, and it
could result from several factors. Though, the energy density
inside the inner radius and the average energy density increase as
$h$ increases for a fixed $r_{\rm{BLR}}$, a majority of the
diffuse radiation cannot make a contribution to the optical depth
$\tau_{\gamma\gamma}$ because of the assumption of
$R_{\rm{\gamma}}=r_{\rm{BLR}}$. Only the diffuse radiation outside
$r_{\rm{BLR}}$ can make a contribution to the optical depth
$\tau_{\rm{\gamma\gamma}}$, and then this results in the
non-monotonous dependence of $\tau_{\rm{\gamma\gamma}}$ on $h$ at
a fixed $r_{\rm{BLR}}$ because the energy density distributions
outside $r_{\rm{BLR}}$ do not monotonously increase with $h$. Only
the fraction of soft photons matching the threshold condition can
make a contribution to the photon-photon absorption, and this
makes a contribution to the non-monotonous dependence on $h$.
These factors together result in this effect that the optical
depth $\tau_{\gamma\gamma}$ does not monotonously increase with
$h$, qualitatively changing the dependence on $h$ with
$E_{\rm{\gamma}}$. If $R_{\rm{\gamma}}=r_{\rm{BLR,in}}$, the
dependence on $h$ at a fixed $r_{\rm{BLR}}$ could be monotonous
because more absorption with $h$ increasing can be contributed by
the diffuse radiation between $r_{\rm{BLR,in}}$ and
$r_{\rm{BLR}}$, and this is justified by the photon-photon
absorption for HFSRQ PKS 0405$-$123 in Figure 8 (see section 4).

In order to study the dependence of $\tau_{\gamma\gamma}$ on the
parameter $R_{\gamma}$, we adopt $r_{\rm{BLR}}=0.43 \/\ \rm{pc}$
and assume the half thickness of BLR $h=0.20 \/\ \rm{pc}$. The
calculated results with four values of $R_{\gamma}$ are presented
in Figure 5. The farther the gamma-ray emitting region from the
central black hole, the less the absorption optical depth of
gamma-rays. When the gamma-ray emitting region is close to the
outer radius of BLR, the two diffuse radiation fields we concerned
are transparent for gamma-rays with energies from 10 to 200 GeV.
If the gamma-ray emitting region is close to the inner radius of
BLR, the two diffuse radiation fields we concerned are not
transparent for gamma-rays between 10 and 200 GeV. This trend of
the absorption optical depth in Figure 5 can be easily explained
from the energy density distributions in Figure 2. The closer the
gamma-ray emitting region to the center of BLR, the larger the
displacement of the escaping gamma-rays and the more the gamma-ray
absorbing photons. The absorption optical depth
$\tau_{\gamma\gamma}$ rapidly decreases as the parameter
$R_{\gamma}$ increases. The photon field energy density sharply
reduces as $R_{\gamma}>r_{\rm{BLR,in}}$, and this makes the
absorption optical depth $\tau_{\gamma\gamma}$ rapidly reduce.

We take four values of $r_{\rm{BLR}}$ to study the dependence of
the absorption optical depth $\tau_{\gamma\gamma}$ on the
parameter $r_{\rm{BLR}}$. In the calculations, it is supposed that
$R_{\gamma}=r_{\rm{BLR}}$ and $h=0.20 \/\ \rm{pc}$. The calculated
results are presented in Figure 6. It can be seen in Figure 6 that
the optical depth $\tau_{\gamma\gamma}$ slowly reduces as the
parameter $r_{\rm{BLR}}$ increases. This trend of the absorption
optical depth can be also easily explained from the energy density
distributions in Figure 2. The increasing of BLR scale
$r_{\rm{BLR}}$ makes the inner and outer radii of BLR become
larger, and this results in the decreasing of the energy density
of radiation fields we concerned, and then the decreasing of the
optical depth of photon-photon absorption.

\section{THE PHOTON-PHOTON ABSORPTION OPTICAL DEPTH FOR FSRQs}

According to Ghisellini et al. (1998), for FSRQs, the Lorentz
factors have the typical value of $\Gamma\sim 15$, the magnetic
energy densities are on the order of $U^{'}_{\rm{B}}\sim 0.1 \/\
\rm{erg \/\ cm^{-3}}$ measured in the blob comoving frame, and the
Compton dominance of $L_{\rm{EC}}/L_{\rm{Syn}}$ is $\sim 20$,
where $L_{\rm{EC}}$ is the EC luminosity and $L_{\rm{Syn}}$ the
synchrotron luminosity. It is suggested that there are relations
of $L_{\rm{EC}}\gg L_{\rm{SSC}}$ and
$L_{\rm{EC}}/L_{\rm{Syn}}\simeq 20$ for FSRQs on average, where
$L_{\rm{SSC}}$ is the SSC luminosity (Georganopoulos et al. 2001).
This Compton dominance $L_{\rm{EC}}/L_{\rm{Syn}}\simeq 20$ is
consistent with the result of Ghisellini et al. (1998). There is a
relationship of $L_{\rm{EC}}/L_{\rm{Syn}}\simeq
U^{'}_{\rm{ph}}/U^{'}_{\rm{B}}=U_{\rm{ph}}/U_{\rm{B}}$ between the
ratio of EC to synchrotron luminosity and the ratio of external
radiation energy density $U_{\rm{ph}}$ to magnetic energy density
$U_{\rm{B}}$ for powerful blazars. The external radiation energy
densities of FSRQs have a typical value of $U_{\rm{ph}}\simeq
U_{\rm{B}}L_{\rm{EC}}/L_{\rm{Syn}}\simeq
\Gamma^{-2}U^{'}_{\rm{B}}L_{\rm{EC}}/L_{\rm{Syn}}\sim 10^{-2.1}
\/\ \rm{erg\/\ cm^{-3}}$. Assuming the ratio of
$\tau_{\rm{BLR}}/f_{\rm{cov}}$ on the order of unity, the
calculations indicate that the average of the external radiation
energy densities within the outer radius $r_{\rm{BLR,out}}$ is on
the order of $\overline{U}_{\rm{ph}}\sim 10^{-2.1} \/\ \rm{erg \/\
cm^{-3}}$ as $r_{\rm{BLR,in}}\sim 0.2 \/\ \rm{pc}$ and
$r_{\rm{BLR,out}}\sim 0.6 \/\ \rm{pc}$. The median of the probable
positions of inner and outer radii is around $0.4 \/\ \rm{pc}$,
which is in well agreement with $r_{\rm{BLR}}\simeq 0.43 \/\
\rm{pc}$ derived from equation (23) and the typical value of
$L_{\rm{BLR}}=10^{45.35} \/\ \rm{erg \/\ s^{-1}}$ for FSRQs. This
agreement implies it reasonable that we adopt the value of
$r_{\rm{BLR}}$ derived from equation (23) as the median of
$r_{\rm{BLR,in}}$ and $r_{\rm{BLR,out}}$ for FSRQs. The typical
values of $L_{\rm{BLR}}=10^{45.35} \/\ \rm{erg \/\ s^{-1}}$ and
$U_{\rm{ph}}\sim 10^{-2.1} \/\ \rm{erg\/\ cm^{-3}}$ for FSRQs
limit the probable positions of inner and outer radii of BLR
around $0.2 \/\ \rm{pc}$ and $0.6 \/\ \rm{pc}$, respectively, and
also confine the half thickness of BLR about $h\sim 0.20 \/\
\rm{pc}$. Then, we adopt $r_{\rm{BLR}}\simeq 0.43 \/\ \rm{pc}$ and
$h\sim 0.20 \/\ \rm{pc}$ to estimate the photon-photon absorption
optical depth of gamma-rays with energies from 10 to 200 GeV for
FSRQs.

However, the gamma-ray emitting regions are still a controversial
issue in the studies on blazars. It has been suggested that
gamma-rays are produced within the BLR, and the gamma-ray emitting
radius $R_{\gamma}$ ranges roughly between 0.03 and 0.3 pc
(Ghisellini \& Madau 1996). The radiative plasma in the
relativistic jets of the powerful blazars are within the BLR
cavities (Georganopoulos et al. 2001), and then there are
$R_{\rm{\gamma}}\lesssim r_{\rm{BLR,in}}$ for FSRQs with the
typical value of $L_{\rm{BLR}}=10^{45.35}\/\ \rm{erg\/\ s^{-1}}$.
Based on the dependence of $\tau_{\gamma\gamma}$ on the parameter
$R_{\gamma}$ discussed in previous section, the closer the
gamma-ray emitting regions to the central black holes, the larger
the optical depth of photon-photon absorption, and then we adopt
$R_{\rm{\gamma}}\sim r_{\rm{BLR,in}}$ to estimate the
$\gamma-\gamma$ absorption optical depth of gamma-rays with
energies from 10 to 200 GeV. Assuming for FSRQs that
$R_{\rm{\gamma}}\sim r_{\rm{BLR,in}}$, $h\sim 0.20 \/\ \rm{pc}$
and $L_{\rm{BLR}}=10^{45.35} \/\ \rm{erg \/\ s^{-1}}$
corresponding to $r_{\rm{BLR}}\simeq 0.43 \/\ \rm{pc}$, we
estimate the photon-photon absorption optical depth of gamma-rays
between 10 and 200 GeV. The line emission is completely
transparent to gamma-rays below 30 GeV (see Fig. 7), because the
threshold condition of photon-photon pair creation process cannot
be satisfied between these gamma-rays below 30 GeV and the
emission lines listed in Figure 3. However, the diluted blackbody
radiation is not transparent to these gamma-rays below 30 GeV (see
Fig. 7). On the other hand, the emission lines contribute much
more than the diluted blackbody radiation to the absorption for
gamma-rays with energies from $\sim$50 to 200 GeV. Even if the
diluted blackbody radiation is not considered in the calculations,
the line emission is not transparent to these gamma-rays with
energies from $\sim$50 to 200 GeV (see Fig. 7). Then, gamma-rays
with energies from 10 to 200 GeV cannot escape from the emission
line and diluted blackbody radiation fields of the BLRs in FSRQs
as long as $R_{\rm{\gamma}}\la r_{\rm{BLR,in}}$ (see Fig. 7). In
this section, the redshift effect is not considered for gamma-rays
with energies from 10 to 200 GeV in the observed spectrum, and
this effect will be considered for individual sources in the
following section.

\section{APPLICATION TO INDIVIDUAL SOURCES}

The discovery of a large population of X-ray strong FSRQs (labeled
HFSRQs by Perlman et al. 1998) gives a fundamental change in our
perception of the broadband spectra of FSRQs. HFSRQs have strong
broad emission lines, flat hard X-ray spectra similar to classical
FSRQs, but high synchrotron peak frequency and steep soft X-ray
spectra similar to intermediate or high synchrotron peak frequency
BL Lac objects, and are classified as a new subclass of blazars
(e.g., Georganopoulos 2000; Padovani et al. 2003). An example of
such an object is RGB J1629+4008 with a synchrotron peak $\sim
2\times 10^{16}\/\ \rm{Hz}$ and a predicted flux $\nu
F_{\rm{\nu}}$ of nearly $10^{-10}\/\ \rm{erg\/\ s^{-1}\/\
cm^{-2}}$ at $10^{25}\/\ \rm{Hz}$ (Padovani et al. 2002).
Therefore, HFSRQs are likely to be GeV--TeV gamma-ray emitters
with the Compton peak energies possibly from multi-ten to
multi-hundred GeV. A powerful HFSRQ with $L_{\rm{BLR}}$ obtained
in literature is PKS 0405$-$123, and it has a BLR luminosity
$L_{\rm{BLR}}=10^{45.67}\/\ \rm{erg\/\ s^{-1}}$ and redshift
$z=0.574$ (e.g., Georganopoulos 2000). Assuming that HFSRQs have
the same properties of BLRs as the classical FSRQs, we estimate
the optical depth of photon-photon absorption for the
strong-lined, powerful HFSRQ PKS 0405$-$123. In the calculations,
we assume $R_{\gamma}\sim r_{\rm{BLR,in}}$ and
$\tau_{\rm{BLR}}/f_{\rm{cov}}=1$, and take three values of $h$,
because it is unclear whether the parameter $h$ equals $\sim 0.20
\/\ \rm{pc}$ for PKS 0405$-$123. From Figure 8 we see that the two
diffuse radiation fields of BLR are not transparent to gamma-rays
with energies from 10 to 200 GeV in the observed spectrum for this
powerful HFSRQ as long as $R_{\rm{\gamma}}\la r_{\rm{BLR,in}}$. If
$\it GLAST$ detect gamma-rays with energies from 10 to 200 GeV
emitted by PKS 0405$-$123, either the gamma-ray emitting regions
are not within the BLR cavity or the model used and the relevant
assumptions made in this paper are incorrect.

The classical FSRQ 3C 279, $z=0.536$, is one of the brightest
extragalactic objects in the gamma-ray sky, and has a BLR
luminosity $L_{\rm{BLR}}=10^{44.41}\/\ \rm{erg\/\ s^{-1}}$ (Cao \&
Jiang 1999). It was detected by the EGRET in the 0.1--10 GeV
energy domain, and its spectrum above $\sim$ 100 MeV can be fitted
by a power law and do not show any signature of gamma-ray
absorption by pair production up to $\sim$ 10 GeV (Fichtel et al.
1994; von Montigny et al. 1995). We calculate the optical depth to
photon-photon absorption for 3C 279 by adopting
$r_{\rm{BLR,in}}=0.1 \/\ \rm{pc}$ and $r_{\rm{BLR,out}}=0.4 \/\
\rm{pc}$ from Hartman et al. (2001) and assuming
$R_{\rm{\gamma}}\sim r_{\rm{BLR,in}}$. The photon-photon
absorption optical depth is presented by the dashed curve in
Figure 8. It can be seen in Figure 8 that the diffuse radiation
fields are not transparent to gamma-rays with energies around 10
GeV, and this result is not consistent with the observations
(e.g., Fichtel et al. 1994; von Montigny et al. 1995; Ghisellini
\& Madau 1996; Wehrle et al. 1998). There are two probable factors
that make the optical depth overestimated. The first one is the
assumption of $R_{\rm{\gamma}}\sim r_{\rm{BLR,in}}$ that can lead
to the overestimation of $\tau_{\rm{\gamma\gamma}}$. The second
one is the assumption of $\tau_{\rm{BLR}}/f_{\rm{cov}}=1$ that can
also result in the same consequence as the first one. If
$\tau_{\rm{BLR}}/f_{\rm{cov}}=1/3$, the photon-photon absorption
optical depth of these gamma-rays around 10 GeV is less than
unity, and the two diffuse radiation fields are optically thick to
these gamma-rays in the interval between multi-ten and 200 GeV.
$\it GLAST$ observations in the future could limit strong
constraints on the position of the gamma-ray emitting region
relative to the inner radius of BLR and the BLR properties for 3C
279.

\section{DISCUSSIONS AND CONCLUSIONS}
It can be seen from equation (12) that the emissivity of emission
lines is unrelated to the covering factor, and then the relevant
absorption for gamma-rays is also unrelated with the value of
$f_{\rm{cov}}$. The gamma-ray absorption contributed by the
diluted blackbody radiation is in proportion to the ratio
$\tau_{\rm{BLR}}/f_{\rm{cov}}$ (see eq. 13). There is a large
uncertainty in the BLR covering factor, and the uncertainty could
affect our results. Early estimates of this quantity indicated a
covering factor $f_{\rm{cov}}\sim 5-10\%$, while recent
observations indicated instead $f_{\rm{cov}}\sim 30\%$ (e.g.,
Maiolino et al. 2001). Estimates of the BLR covering factors need
to measure UV continuum emission from accretion disks. In the case
of blazars, these UV continuum emission are swamped by the beaming
components from relativistic jets, and this makes it difficult to
calculate the BLR covering factors for blazars. The way out to
roughly estimate $f_{\rm{cov}}$ is to know the BLR luminosity
$L_{\rm{BLR}}$ of blazars, in which the thermal emission is
clearly dominating the synchrotron ones. D'Elia et al. (2003)
found only two of these blazars in literatures: 3C 273 and PKS
2149-306. They obtained $f_{\rm{cov}}\sim 7\%$ for the first
source, while $f_{\rm{cov}}\sim 10\%$ for the second one. D'Elia
et al. (2003) adopted the BLR covering factor $f_{\rm{cov}}\sim
10\%$ as the typical value of $f_{\rm{cov}}$ for their blazar
sample. Donea \& Protheroe (2003) suggested $f_{\rm{cov}}\sim 3\%$
for the spherical distribution of the BLR clouds in quasars. The
issue of the Thomson optical depth of BLRs has been rarely studied
and has gotten nowhere for blazars. Blandford \& Levison (1995)
adopted the Thomson optical depth of BLR as
$\tau_{\rm{BLR}}=0.01$. Then the ratio
$\tau_{\rm{BLR}}/f_{\rm{cov}}$ of the Thomson optical depth to the
covering factor is likely to be typically of
$\tau_{\rm{BLR}}/f_{\rm{cov}}\sim 1/3$ for the spherical
distribution of the BLR clouds in blazars. This value of
$\tau_{\rm{BLR}}/f_{\rm{cov}}\sim 1/3$ is on the same order of
magnitude as $\tau_{\rm{BLR}}/f_{\rm{cov}}=1$, assumed in the
previous sections.

As stated in section 3, the external radiation energy densities of
FSRQs are on the order of $U_{\rm{ph}}\sim 10^{-2.1} \/\
\rm{erg\/\ cm^{-3}}$, and this typical value limits constraints on
the probable positions of inner and outer radii of BLR or the half
thickness of spherical shell of BLR. The probable positions of the
inner and outer radii are around $r_{\rm{BLR,in}}\sim 0.2 \/\
\rm{pc}$ and $r_{\rm{BLR,out}}\sim 0.6 \/\ \rm{pc}$, respectively.
The median of this probable inner and outer radii is around $0.4
\/\ \rm{pc}$, which is consistent with the radius
$r_{\rm{BLR}}\simeq 0.43 \/\ \rm{pc}$ derived from the
reverberation mapping. This agreement implies that the radius
$r_{\rm{BLR}}$ derived from the reverberation mapping probably
denotes the median of $r_{\rm{BLR,in}}$ and $r_{\rm{BLR,out}}$,
and our assumption of $r_{\rm{BLR}}$ in the previous sections is
appropriate. The energy density distribution for the BLR with
$r_{\rm{BLR,in}}\sim 0.2 \/\ \rm{pc}$, $r_{\rm{BLR,out}}\sim 0.6
\/\ \rm{pc}$ and $\tau_{\rm{BLR}}/f_{\rm{cov}}\sim 1/3$ has an
average of energy densities within the outer radius on the order
of $U_{\rm{ph}}\sim 10^{-2.1} \/\ \rm{erg \/\ cm^{-3}}$. The
corresponding absorption optical depth for gamma-rays is presented
by the curve III in Figure 7, from which it can be seen that the
two diffuse radiation fields of the BLRs are not transparent to
these gamma-rays with energies from 10 to 200 GeV.

Because of the relativistic beaming effect, most of gamma-rays for
FSRQs are contained within a radiation cone with a half open angle
of $\Delta \varphi\sim 1/ \Gamma\sim 3.8^{\circ}$. If the central
UV photons directly from accretion disks travel through the
radiation cone, the UV photons can have photon-photon pair
creation processes with gamma-rays within the radiation cone.
Assuming the radius of the UV radiation region is around
$R_{D}\sim 100 R_{\rm{s}}\sim 0.001 \/\ \rm{pc}$
($R_{\rm{s}}=2GM/c^2$ and $M=10^8 M_{\rm{\odot}}$), and the
gamma-ray emitting region $R_{\rm{\gamma}}\sim 0.1 \/\ \rm{pc}$,
the angle between the jet direction and the travelling direction
of UV photons at $R_{\rm{\gamma}}$ is $\sim \arcsin
R_{\rm{D}}/R_{\rm{\gamma}}\sim 0.57^{\circ}$. Then the photons
within the radiation cone have the collision angles of $\theta
\lesssim 4.4^{\circ}$. For the UV photons at the frequency $\nu
\simeq10^{16.5} \/\ \rm{Hz}$ with energies of $\varepsilon_2\simeq
2.56\times 10^{-4}$ and these gamma-rays of $\varepsilon_1\simeq
3.91\times 10^{5}$ corresponding to energies around 200 GeV, the
left of equation (3) has a upper limit of $\lesssim 0.15$, which
is less than the value at the right of equation (3), and then the
two kinds of photons cannot be absorbed by the photon-photon pair
creation processes. Therefore, the central UV radiation have
negligible contributions to the absorption for gamma-rays relative
to the diffuse radiation from the BLRs.

The energies carried by the annihilated gamma-rays would be
deposited in the lower frequency radiation field by the
synchrotron and inverse Compton processes of the electron-positron
pairs and the pair cascade and pair annihilation processes (see
Protheroe \& Stanev 1993; Saug\'e \& Henri 2004; Zdziarski \&
Coppi 1991). The intense creation of pairs would produce a strong
radiation at low energy X-rays or at GeV energy (Protheroe \&
Stanev 1993; Saug\'e \& Henri 2004; Zdziarski \& Coppi 1991). The
electron-positron pair cascade can cause the soft X-ray excesses
(Zdziarski \& Coppi 1991). The produced electron-positron pairs
could make a difference at lower energies, around 1 GeV, where the
pileup of photons below the absorption feature results in a
significant flattening in the observed spectrum relative to the
emission spectrum (Protheroe \& Stanev 1993). Then the
electron-positron pairs produced by the photon-photon annihilation
processes in FSRQs could result in the similar observational
effects mentioned above at around 1 GeV gamma-rays or low energy
X-rays.

In summary, we used a BLR model to study the photon-photon
absorption optical depth for gamma-rays with energies from 10 to
200 GeV in the observed spectrum. There are four key parameters
for the BLR model: the gamma-ray emitting radius $R_{\gamma}$, the
BLR luminosity $L_{\rm{BLR}}$, the half thickness of BLR $h$ and
the ratio $\tau_{\rm{BLR}}/f_{\rm{cov}}$. The BLR luminosity
$L_{\rm{BLR}}$ determines the distance $r_{\rm{BLR}}$, which
limits the BLR inner and outer radii of $r_{\rm{BLR,in}}$ and
$r_{\rm{BLR,out}}$ if the parameter $h$ is given. We discussed the
dependence of the optical depth $\tau_{\gamma\gamma}$ on the
parameters $R_{\rm{\gamma}}$, $h$ and $r_{\rm{BLR}}$. The diffuse
external radiation energy density in the gamma-ray emitting region
is typically on the order of $U_{\rm{ph}}\sim 10^{-2} \/\ \rm{erg
\/\ cm^{-3}}$ for FSRQs, and the ratio
$\tau_{\rm{BLR}}/f_{\rm{cov}}$ might be typically on the order of
$\tau_{\rm{BLR}}/f_{\rm{cov}}\sim 1$ for blazars. Then the inner
and outer radii of BLR are required to be around
$r_{\rm{BLR,in}}\sim 0.2 \/\ \rm{pc}$ and $r_{\rm{BLR,out}}\sim
0.6 \/\ \rm{pc}$, respectively. It is impossible for gamma-rays
from 10 to 200 GeV emitted by FSRQs to escape from the diffuse
radiation fields of the BLRs if $R_{\gamma}\sim r_{\rm{BLR,in}}$
(see Fig. 7). If $\it GLAST$ could detect gamma-rays with energies
from 10 to 200 GeV for most of FSRQs, there would be
$R_{\rm{\gamma}}>r_{\rm{BLR,in}}$ for FSRQs on average, otherwise,
$R_{\rm{\gamma}}\lesssim r_{\rm{BLR,in}}$. If
$\tau_{\rm{BLR}}/f_{\rm{cov}}\sim 1/3$ and $R_{\rm{\gamma}}\sim
r_{\rm{BLR,in}}$ for the classical FSRQ 3C 279, the diffuse soft
photon field is not transparent to gamma-rays with energies from
multi-ten to 200 GeV and is transparent to gamma-rays around 10
GeV. $\it GLAST$ observations in the energy domain of multi-ten to
200 GeV could limit constraints on the position of the gamma-ray
emitting region relative to the inner radius of BLR and the BLR
properties for 3C 279. For the powerful HFSRQ PKS 0405$-$123, the
diffuse photon field is not transparent to gamma-rays with
energies from 10 to 200 GeV in the observed spectrum. $\it GLAST$
observations in the future could limit constraints on the
gamma-ray emitting region and the BLR in this HFSRQ. Our results
are model dependent, especially dependent on the gamma-ray
emitting radius $R_{\rm{\gamma}}$ and the inner radius
$r_{\rm{BLR,in}}$. To determine the relationship of the inner
radius $r_{\rm{BLR,in}}$ with the radius $r_{\rm{BLR}}$ derived
from the reverberation mapping needs more theoretical and
observational researches. $\it GLAST$ observations in the future
can give more observational constraints on the gamma-ray emitting
regions and the BLRs for the powerful blazars.

\acknowledgements We are grateful to the anonymous referee for
his/her constructive comments and suggestions that helped to
significantly improve this paper. This work is supported by
National Natural Science Foundation of China (Grant 10443003,
10573030, and 10533050) and Natural Science Foundation of Yunnan
Province (Grant 2003A0025Q). J.M.B. thanks support of the Bai Ren
Ji Hua project of the CAS, China.

\clearpage
\begin{figure}
\includegraphics[angle=-90,scale=.5]{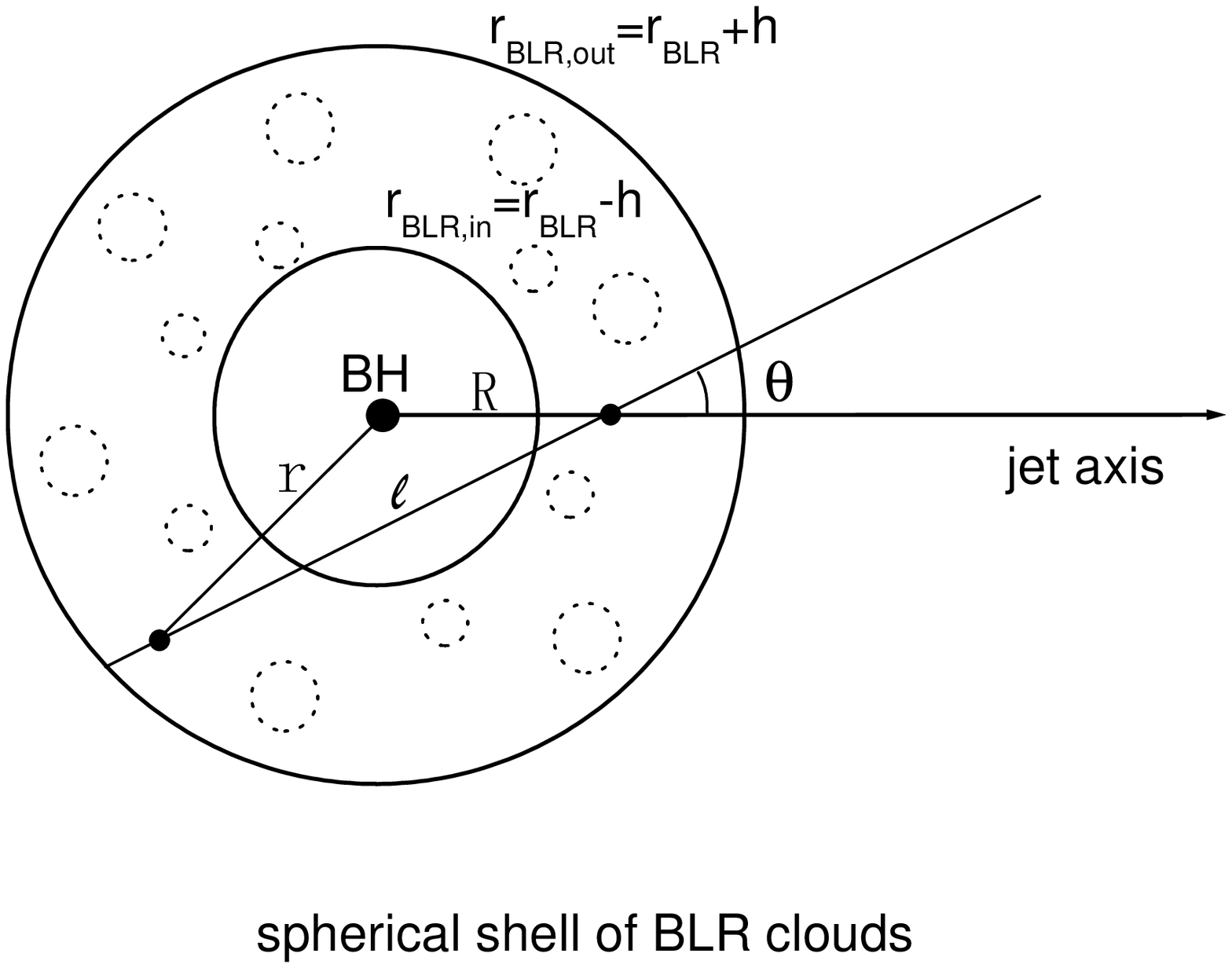}
 \caption{Sketch of axial cross section illustrating the geometry used for computation of the intensity
 of the BLR radiation and similar to Fig. 4 of Donea \& Protheroe (2003).}
\label{fig1}
\end{figure}

\begin{figure}
\includegraphics[angle=-90,scale=.55]{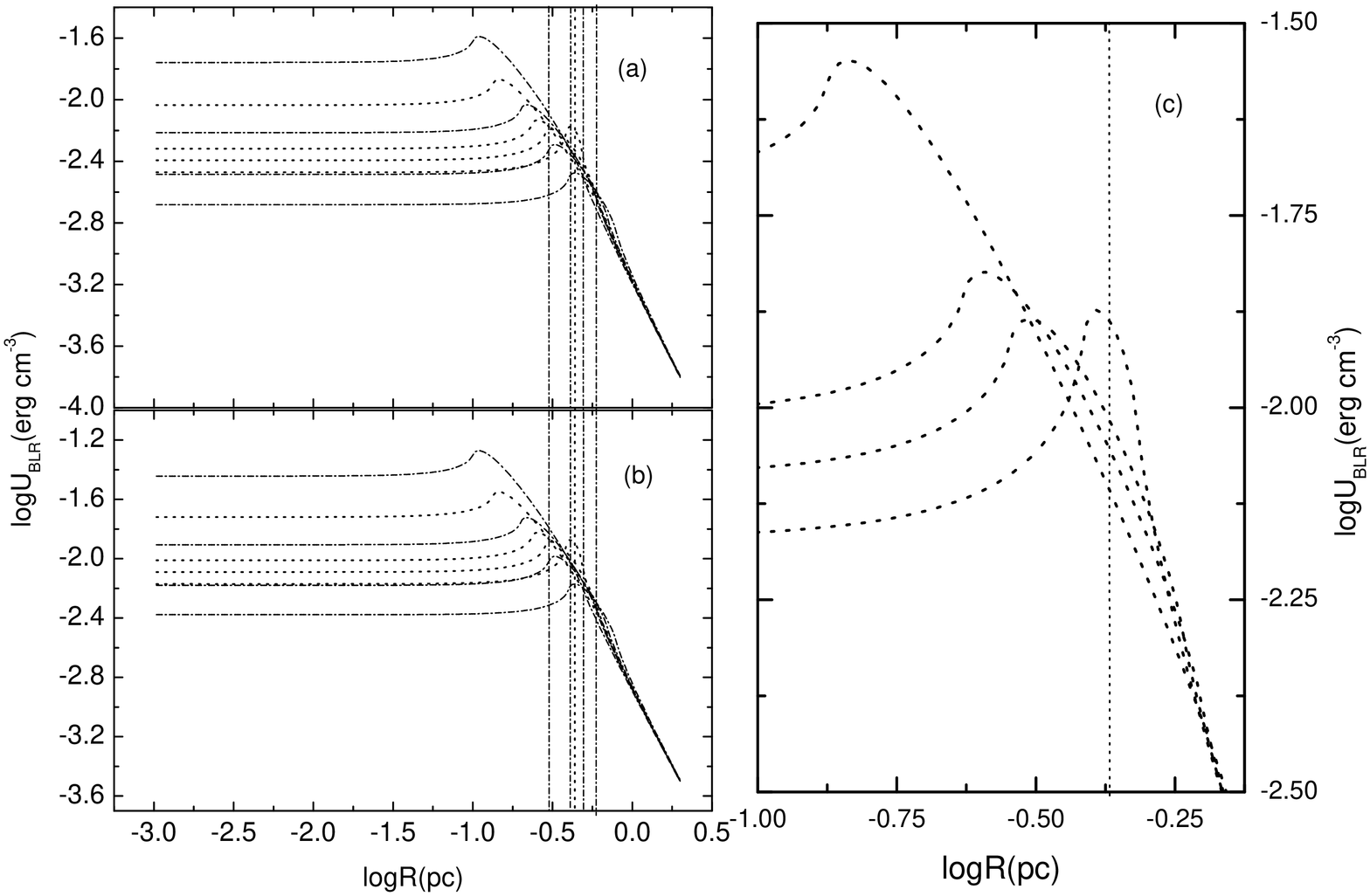}
 \caption{The radial dependence of the energy density on the parameters $r_{\rm{BLR}}$ and $h$.
 The dotted curves in plots are calculated by adopting $L_{\rm{BLR}}=10^{45.35}\/\ \rm{erg\/\ s^{-1}}$,
 $r_{\rm{BLR}}=0.43 \/\ \rm{pc}$, and $\tau_{\rm{BLR}}/f_{\rm{cov}}=1$. From the bottom up, the four dotted curves
 have the parameter $h$ at 0.05, 0.15, 0.20, and $0.30\/\ \rm{pc}$.
 The dash-dotted curves are calculated by adopting $L_{\rm{BLR}}=10^{45.35}\/\ \rm{erg\/\ s^{-1}}$, $h=0.20\/\ \rm{pc}$,
 and $\tau_{\rm{BLR}}/f_{\rm{cov}}=1$. From the top down, the four
 dash-dotted curves have the parameter $r_{\rm{BLR}}$ at 0.30, 0.40, 0.50, and $0.60\/\ \rm{pc}$.
 The plot (a) presents the energy density distributions for the emission lines, and the plot (b) presents the corresponding
 energy density distributions for the emission lines and continuum radiation. The plot (c) is the zoomed parts for the
 four dotted curves in the plot (b). The dotted vertical lines correspond to $r_{\rm{BLR}}=0.43 \/\ \rm{pc}$, and the
 dash-dotted vertical lines correspond to $r_{\rm{BLR}}$ at 0.30, 0.40, 0.50, and $0.60\/\ \rm{pc}$.} \label{fig2}
\end{figure}

\begin{figure}
\includegraphics[angle=-90,scale=.50]{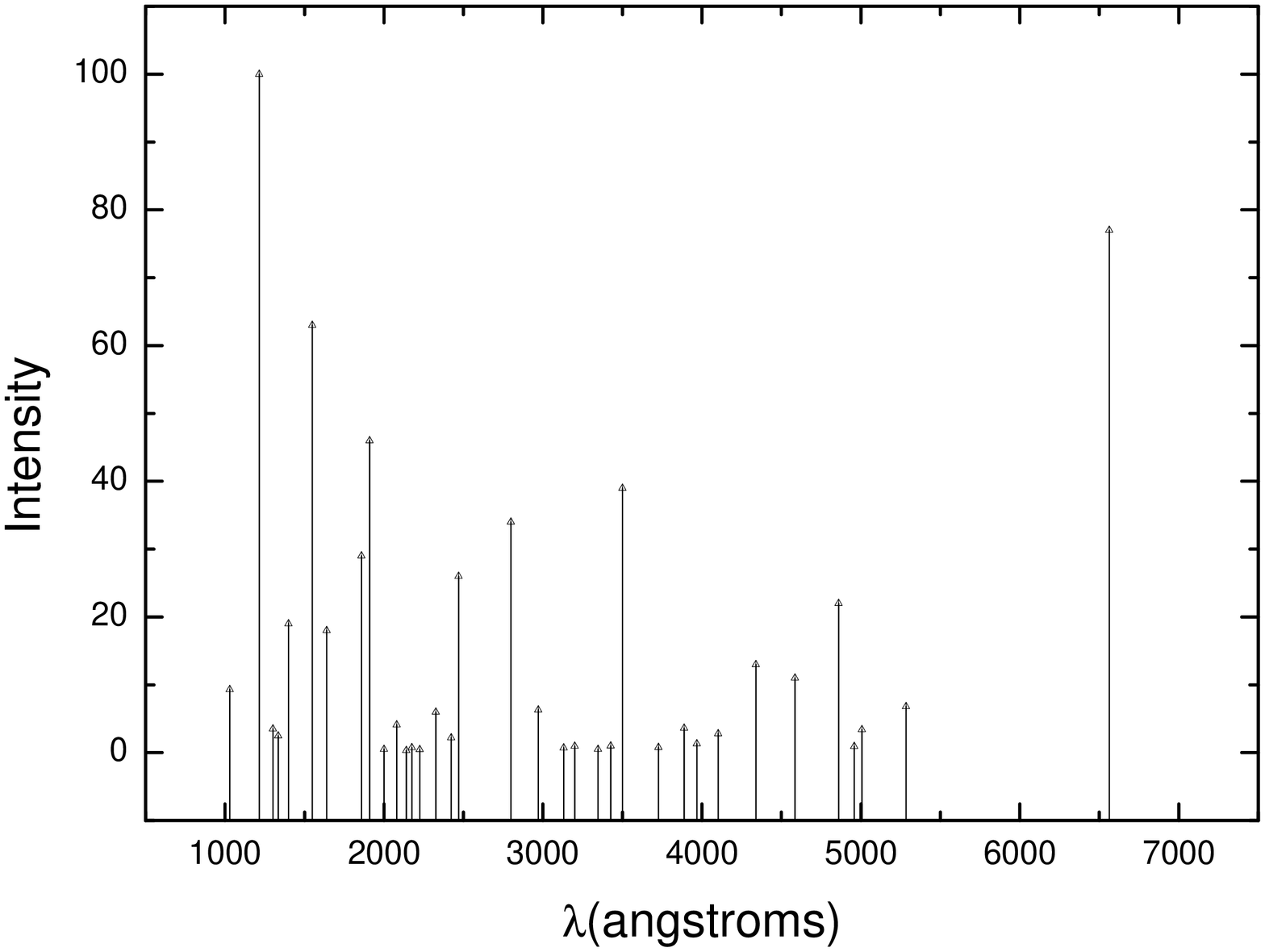}
 \caption{The relative intensity of broad emission lines adopted by us.}
\label{fig3}
\end{figure}

\begin{figure}
\includegraphics[angle=-90,scale=.55]{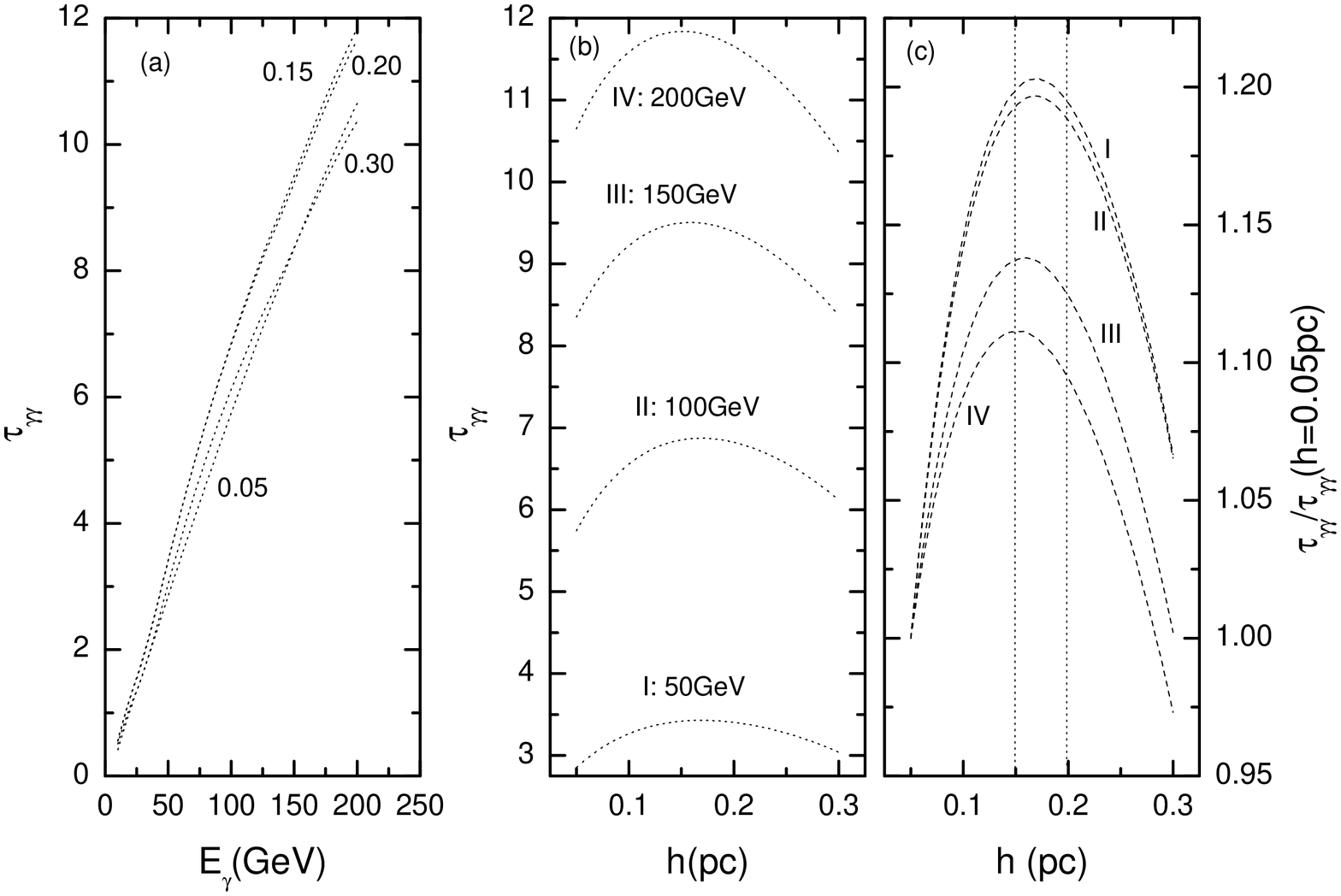}
 \caption{The dependence of $\tau_{\gamma\gamma}$ on the parameter $h$. In the calculations, we used
 $r_{\rm{BLR}}=0.43\/\ \rm{pc}$ and assumed $R_{\rm{\gamma}}=r_{\rm{BLR}}$ and $\tau_{\rm{BLR}}/f_{\rm{cov}}=1$.
 The plot (a) is the photon-photon optical depth for four values of $h$, and the numbers attached to the four curves
 give $h$ in $\rm{pc}$. The plot (b) is the dependence of $\tau_{\gamma\gamma}$ on $h$ for four values of $E_{\rm{\gamma}}$.
 The plot (c) is the normalized $\tau_{\gamma\gamma}$ corresponding to the four cases in the plot (b).} \label{fig4}
\end{figure}

\begin{figure}
\includegraphics[angle=-90,scale=.50]{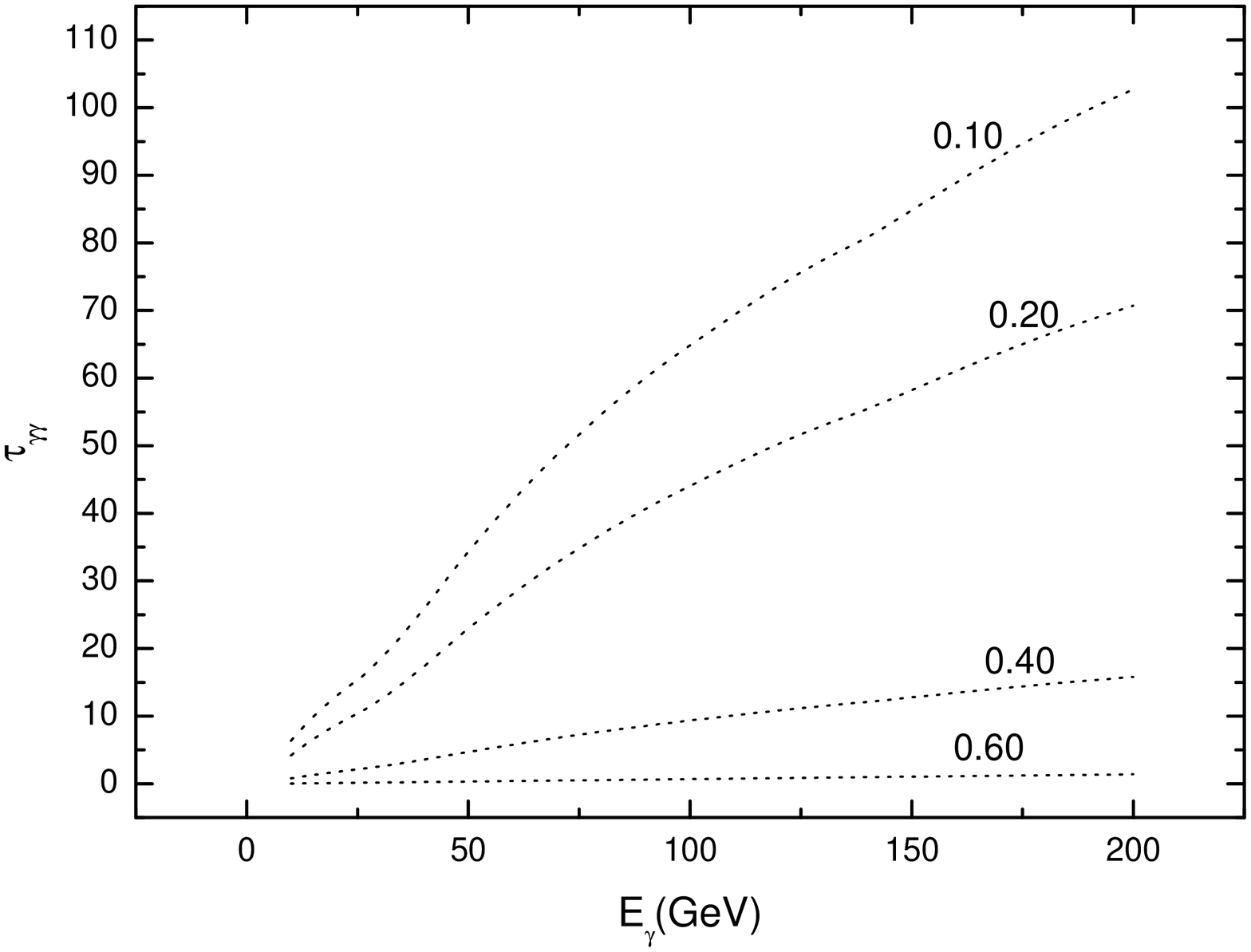}
 \caption{The dependence of $\tau_{\gamma\gamma}$ on the parameter $R_{\gamma}$, and numbers
 attached to the curves give $R_{\gamma}$ in $\rm{pc}$. In the calculations, $r_{\rm{BLR}}=0.43 \/\ \rm{pc}$,
 $h=0.20 \/\ \rm{pc}$ and $\tau_{\rm{BLR}}/f_{\rm{cov}}=1$ are assumed.} \label{fig5}
\end{figure}

\begin{figure}
\includegraphics[angle=-90,scale=.50]{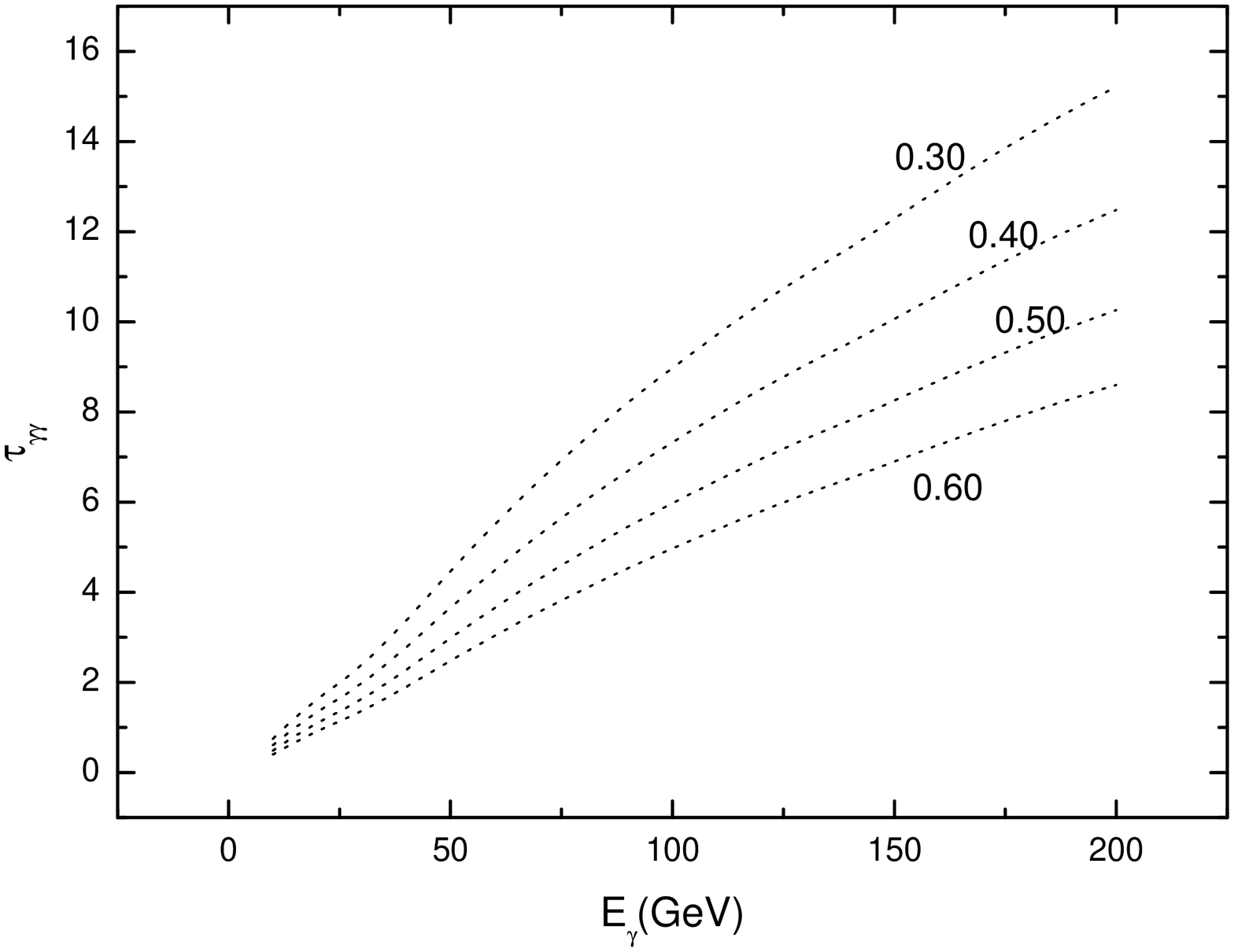}
 \caption{The dependence of $\tau_{\gamma\gamma}$ on the parameter $r_{\rm{BLR}}$, and numbers
 attached to the curves give $r_{\rm{BLR}}$ in $\rm{pc}$. In the calculations, $R_{\rm{\gamma}}=r_{\rm{BLR}}$,
 $h=0.20 \/\ \rm{pc}$, and $\tau_{\rm{BLR}}/f_{\rm{cov}}=1$ are assumed.} \label{fig6}
\end{figure}

\begin{figure}
\includegraphics[angle=-90,scale=.50]{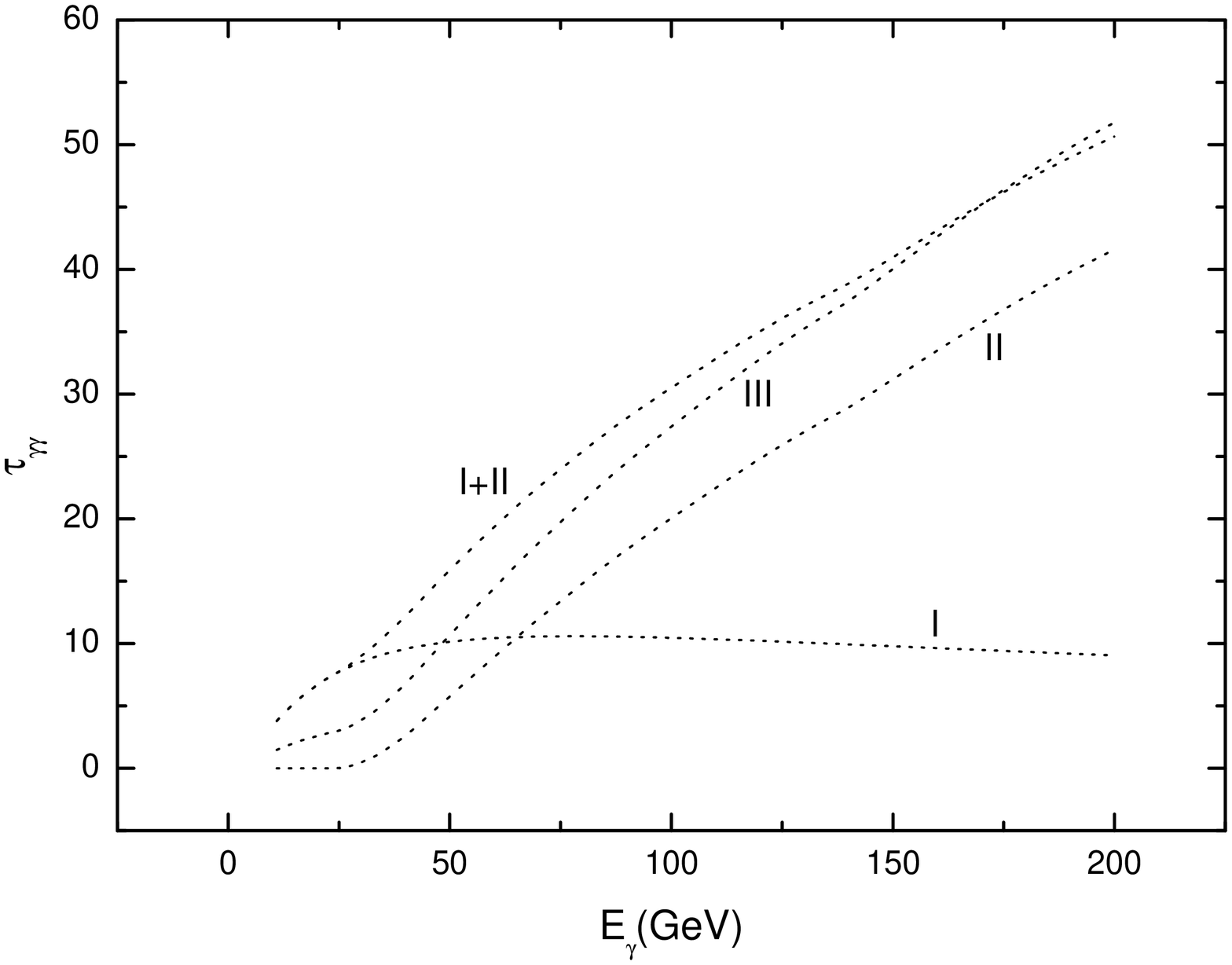}
 \caption{The optical depth of photon-photon absorption for FSRQs. The curve I corresponds to the
 absorption by the diluted blackbody radiation, the curve II the absorption by the emission lines, and the
 curve I+II the sum of the curves I and II. The three curves are calculated by adopting $r_{\rm{BLR}}=0.43 \/\ \rm{pc}$,
 $h=0.20 \/\ \rm{pc}$, $R_{\rm{\gamma}}=r_{\rm{BLR,in}}$, and $\tau_{\rm{BLR}}/f_{\rm{cov}}=1$. The curve III corresponds
  to the optical depth by adopting $r_{\rm{BLR,in}}= 0.2 \/\ \rm{pc}$, $r_{\rm{BLR,out}}= 0.6 \/\ \rm{pc}$,
  $\tau_{\rm{BLR}}/f_{\rm{cov}}= 1/3$, and $R_{\rm{\gamma}}=r_{\rm{BLR,in}}$.}
 \label{fig7}
\end{figure}

\begin{figure}
\includegraphics[angle=-90,scale=.50]{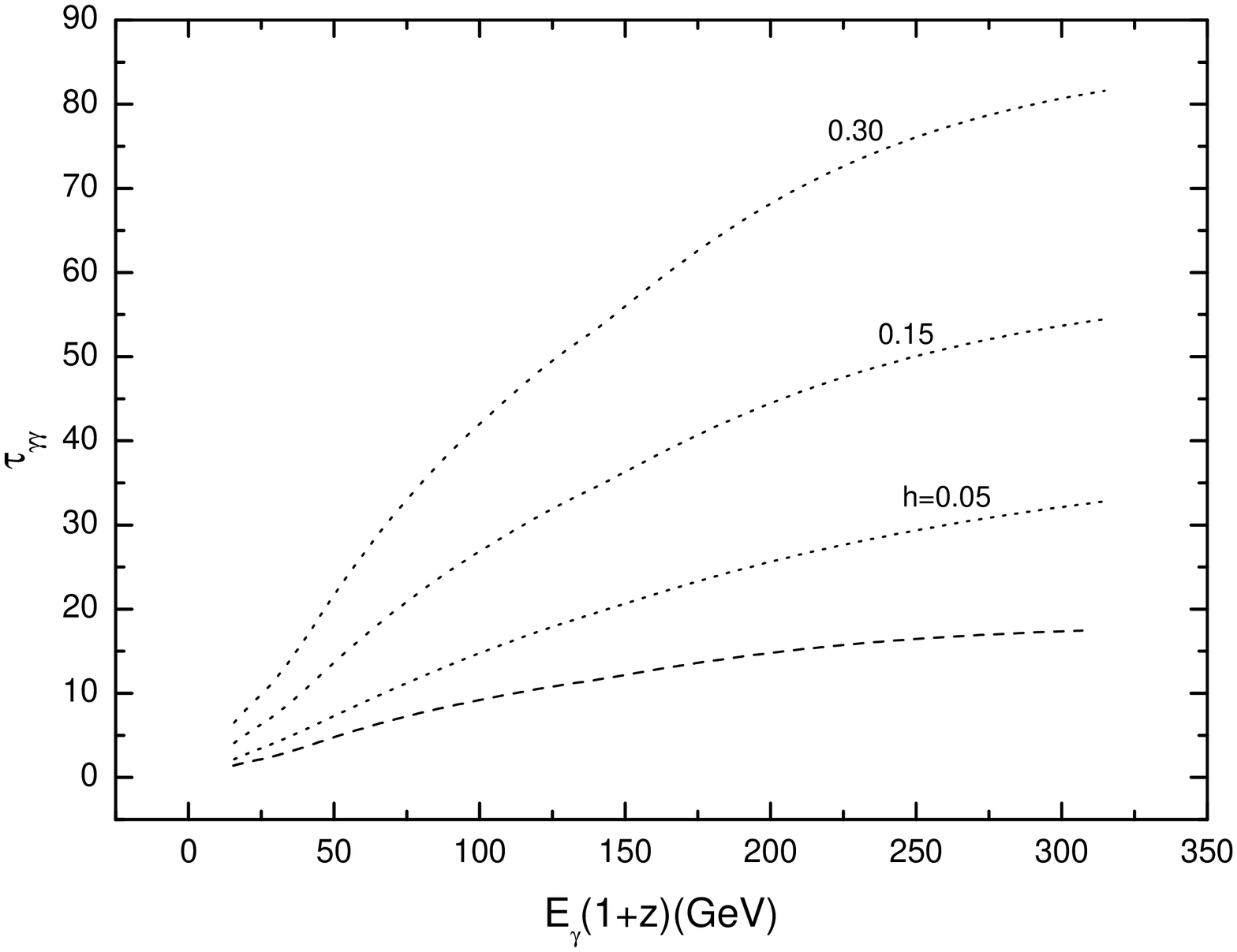}
 \caption{The photon-photon optical depth for HFSRQ PKS 0405$-$123 are presented by the dotted curves, and numbers
 attached to the curves give $h$ in $\rm{pc}$. The dashed curve is the photon-photon optical depth for 3C 279,
 adopting $r_{\rm{BLR,in}}=0.1 \/\ \rm{pc}$ and $r_{\rm{BLR,out}}=0.4 \/\ \rm{pc}$. In the calculations,
 $R_{\rm{\gamma}}=r_{\rm{BLR,in}}$ and $\tau_{\rm{BLR}}/f_{\rm{cov}}=1$ are assumed.}
\label{fig8}
\end{figure}

\end{document}